\documentclass[conference]{IEEEtran}

\let\theorem\relax

\let\lemma\relax
 
\usepackage{float}   
\usepackage{amsmath}
\usepackage{amssymb}
\usepackage{amsthm}
\usepackage{amsfonts}
\usepackage{array}
\usepackage{color}
\usepackage{enumerate}
\usepackage{caption}
\usepackage{subcaption}
\usepackage{framed}
\usepackage{footnote}
\usepackage{boxedminipage}
\usepackage{balance}
\usepackage{epsfig}
\usepackage{textcomp}

\usepackage{latexsym}
\usepackage{graphicx}

\usepackage{multirow}
\usepackage{thmtools}
\def\BibTeX{{\rm B\kern-.05em{\sc i\kern-.025em b}\kern-.08em
    T\kern-.1667em\lower.7ex\hbox{E}\kern-.125emX}}

\newcommand{\KeyGen}{\mathsf{KeyGen}}
\newcommand{\InitAcc}{\mathsf{InitAcc}}
\newcommand{\Add}{\mathsf{Add}}
\newcommand{\Delete}{\mathsf{Delete}}

\newcommand{\VerifyMem}{\mathsf{VerifyMem}}
\newcommand{\VerifyNonMem}{\mathsf{VerifyNonMem}}
\newcommand{\CheckUpdate}{\mathsf{CheckUpdate}}

\newcommand{\Constructor}{\mathsf{Constructor}}
\newcommand{\Register}{\mathsf{Register}}
\newcommand{\Revoke}{\mathsf{Revoke}}
\newcommand{\RetrieveState}{\mathsf{RetrieveState}}

\newcommand{\Map}{\mathsf{Map}}

\newcommand{\VerifySig}{\mathsf{VerifySig}}

\newcommand{\fail}{\mathsf{fail}}

\newcommand{\ignore}[1]{}

\begin{document}

\title{On the Practicality of a Smart Contract PKI}

\author{\IEEEauthorblockN{Christos Patsonakis\IEEEauthorrefmark{1}, Katerina Samari\IEEEauthorrefmark{1}, Aggelos Kiayias\IEEEauthorrefmark{2} and Mema Roussopoulos\IEEEauthorrefmark{1}}
\IEEEauthorblockA{\IEEEauthorrefmark{1}University of Athens, Greece,
 \{c.patswnakis,ksamari,mema\}@di.uoa.gr}
\IEEEauthorblockA{\IEEEauthorrefmark{2}University of Edinburgh and IOHK, UK,
 Aggelos.Kiayias@ed.ac.uk}}
\maketitle

\begin{abstract}

Public key infrastructures (PKIs) are one of the main building blocks for securing
communications over the Internet. Currently, PKIs are under the control of centralized
authorities, which is problematic as evidenced by numerous incidents where they have
been compromised. The distributed, fault tolerant log of transactions provided
by blockchains and more recently, smart contract platforms, constitutes a powerful tool
for the decentralization of PKIs. To verify the validity of identity records,
blockchain-based identity systems store on chain either all identity records, or, a
small (or even constant)
sized amount of data to verify identity records stored off chain. However, as most of
these systems have never been implemented, there is little
information regarding the practical implications of each design's tradeoffs.

In this work,
we first implement and evaluate the only provably secure, smart contract based
PKI of \cite{ourpaper} on top of Ethereum. This construction incurs constant-sized storage
at
the expense of computational complexity. To explore this tradeoff, we propose and implement
a second construction which, eliminates the need for trusted setup, 
preserves the security properties of \cite{ourpaper} and, as illustrated through our
evaluation, is the only
version with constant-sized state that can be deployed on the live chain of
Ethereum. Furthermore, we compare these two systems with the simple approach of most
prior works, e.g., the Ethereum Name Service, where all identity records are stored on the
smart contract's state, to illustrate several shortcomings of Ethereum and its cost model. We
propose several  modifications for fine tuning the model, which would be  useful to be considered for any  smart contract platform like Ethereum
so that it reaches its full potential to support arbitrary distributed applications.

\end{abstract}

\section{Introduction}
\label{sec:introduction}

Contrary to its original, clean design principles, the Internet 
today is not completely decentralized.
For example, Domain Name System (DNS) and Public Key Infrastructure (PKI) services, provide the most critical 
building blocks for facilitating and securing communications over the
Internet.  These systems manage mappings between identity names and values 
(i.e., an IP address for DNS, or a public-key for PKIs).  Unfortunately,
these critical systems are under the control of centralized, 
remote parties that must
be trusted to function correctly.  
This is problematic as evidenced by the numerous incidents in, e.g., centralized
PKIs, where
certification authorities (CAs) have been compromised
(e.g.,~\cite{diginotarcomp,symantec,trustwave}). 

Since the advent of Bitcoin, blockchains
show promise for building systems that are completely 
distributed with no trusted parties.  
Blockchains solve the well studied
problem of distributed consensus (\cite{lamportbyzconsensus}) in an open networking
environment.  They provide a distributed, fault-tolerant, auditable, append-only ledger of transactions.  
As a result of this potential, there
have been calls from the community to ``re-decentralize" the Internet
by leveraging blockchain technologies to
build critical naming and PKI services and, thus, eliminate the
Internet's reliance on centralized entities (e.g.,~\cite{blockstack,namecoin}).

Notable examples of blockchain-based identity management systems are
Namecoin (\cite{namecoin}) and Emercoin (\cite{emercoin}).
These systems employ the blockchain to store, verify and query for records 
pertaining to
identities. However, this approach is inefficient for several reasons.  First, it forces 
clients to download and
maintain an entire copy of the blockchain to verify records.  Second, 
computation and storage
requirements scale linearly with the number of registered records.  Third, it limits the 
system's applicability by excluding important, storage-limited devices, e.g., smartphones. 
Finally, it bloats the blockchain, increasing the size of the state
that miners have to maintain, which may not align well with incentives
of new miners
that wish to sync  and contribute to the blockchain's security.

To deal with these inefficiencies, researchers have recently proposed systems
based on next-generation blockchains, i.e., smart contract platforms
(\cite{certcoin,ourpaper}).  These systems decouple storage of identity records from their verification.
Storage of identity records is handled by a separate,  
authenticated (but potentially unreliable) storage network which
allows for a more compact retrieval of the full history of operations, compared to
downloading and validating the entire blockchain. 
To verify the validity of identity records, they maintain on the
smart contract's state cryptographic accumulators, which are space-efficient data structures that
allow for verifiable membership and non membership queries. 

Until now,
these smart contract based identity management systems have not been implemented
and, thus, there is little practical experience to guide developers 
and to inform design tradeoffs of future systems.
In this paper, we examine the practical issues in building smart contract
based PKI systems.  
We describe our experience of implementing on Ethereum the recently proposed,
state-of-the-art,
smart contract based PKI of Patsonakis et al.~\cite{ourpaper}.  
We choose this system because
it is the only PKI with a security model and proof that it provides the claimed
service while, at the same time, consuming constant-sized state,
as opposed to prior blockchain-based PKIs.  
The system's main building block is a public state, additive, universal accumulator
based on the strong-RSA assumption. This accumulator favors storage overhead
at the expense of computational overhead necessary to achieve constant-sized
state and proofs. To explore the storage versus computational cost tradeoff, we propose
a second smart contract based PKI built on top of the Hash tree-based, universal
accumulator of~\cite{CamachoHKO08}.  Unlike the RSA-based accumulator, the Hash tree-based 
accumulator does not require a trusted setup and, thus, leads to a truly decentralized system
while, at the same time, preserves the same security model and properties of the RSA-based
approach.  We also compare both of these systems with the simple approach of most prior
schemes, e.g., the Ethereum Name Service (\cite{ethens}), where all identity records are stored on the
smart contract's state and we illustrate several shortcomings of Ethereum and its cost model.

In summary, the contributions of this paper are as follows. 
First, we evaluate experimentally all three smart contract based PKI schemes and illustrate the monetary costs of
their operations, as well as those of their building blocks. Our 
results illustrate
that the Hash tree-based construction is the only smart contract based PKI with constant-sized state that 
can be deployed on Ethereum's live
chain, thus, providing the first viable and provably secure, on-blockchain authentication
mechanism. 
Second, we present several shortcomings of Ethereum's current
cost model and the ways in which: i) it affects each version of the PKI system, 
ii) it impedes the
establishment of a standard library of smart contracts, iii) it incentivizes smart contract
developers to adopt several malpractices and, iv) it prices on-blockchain storage of data.
Third, we propose several modifications to Ethereum's cost model, which are
minor and fair, to address the aforementioned issues and others. 
Finally, we identify several problem areas we encountered while developing
on top of Ethereum and make concrete proposals for improving the platform
with the aim of increasing both developer productivity and smart contract
reliability. We argue that these improvements are sensible for any smart contract platform that wishes to support user developed distributed applications.

\section{Related Work}
\label{sec:relatedwork}

There is a large number of prior works on PKIs, in general. Due to lack
of space, we review related work specifically
on decentralized,
blockchain and smart contract based PKIs.

Several works propose systems that employ the blockchain to store,
query and verify the validity of \textbf{each} (identity,public-key) pair, or,
some representation (e.g., hash) of it
(\cite{namecoin,emercoin,ethens,scpki,bbpki,ikp,certrans,ndn}). Blockstack 
(\cite{blockstack}) allows the development of arbitrary state machines
via its \emph{virtualchains}, i.e., it is a smart 
contract platform. Blockstack's BNS is a virtualchain (smart contract)
that implements a distributed DNS. BNS stores the hashes of DNS zone files
(\cite{zonefile}) on Atlas, Blockstack's distributed peer network. BNS clients
verify the validity of each zone file hash stored in Atlas by searching the 
blockchain for a BNS transaction that contains this hash. Thus, BNS incurs
the same inefficiencies of Namecoin and Emercoin, which we discussed
in the Introduction. In contrast, our Hash tree-based PKI follows a different design
principle that stores on the smart contract's state a constant and verifiable
representation of \textbf{all} (identity,public-key) pairs by employing cryptographic accumulators. Clients of
our PKI can verify the validity of records pertaining to identities via these
accumulators and by interacting with a (potentially unreliable) storage network. Thus, in our PKI, there is
no need to linearly search the blockchain. In addition, our design can be
implemented on top of any system that allows the development of smart
contracts. Thus, our Hash tree-based PKI can be even implemented on top of
Blockstack. Nonetheless, we chose to implement
our construction on top of Ethereum because it 
has a more rich and diverse ecosystem of applications. Multisignature
wallets (e.g., \cite{ethmultisig}) and various (non) fungible tokens
(e.g., \cite{erc20tokencontracts}) are just a couple example
applications that can benefit from the standard, on-blockchain
authentication mechanism that our construction provides.

Melara et al.~\cite{coniks} introduce CONIKS, a privacy-preserving decentralized PKI where users can monitor the consistency of their own
(identity,public-key) pairs. While privacy is an important property,
e.g., for chat applications, it is not a requirement for traditional
PKIs. For instance, in the web-PKI paradigm, the (identity,public-key)
pairs of participants are public. CONIKS's 
operation is based on ``identity providers'', i.e.,
centralized entities that sign authenticated bindings and
appropriately transform identity names for privacy purposes.
CONIKS assumes the existence of a separate PKI to distribute the public
keys of identity providers. Thus, it does not constitute a standalone
PKI service, whilst our Hash tree-based PKI does. More importantly,
CONIKS, lacks a formal proof of its security and privacy guarantees.
This is also the case for systems derived from CONIKS, i.e., 
EthIKS (\cite{ethiks}), Catena (\cite{catena}) and Conifer
(\cite{conifer}), which implement CONIKS on
top of Ethereum and Bitcoin. Certcoin (\cite{certcoin}) is a
blockchain-based PKI proposal that employs cryptographic accumulators but has
a number of inefficiencies, e.g., it recomputes, from scratch, accumulator
values during \textbf{each} revocation. Furthermore, Certcoin has no security
model for the PKI it implements nor a proof that it provides the claimed
service (\cite{ourpaper}).

Formal proofs of
security are essential for critical security
infrastructures. The Hash tree-based PKI we present here
is built on top of the hash tree, universal accumulator
of Camacho et al.~\cite{CamachoHKO08}, which is a public-state,
additive, universal accumulator.  Thus, it conforms to the design of
the smart contract based PKI of Patsonakis et al.~\cite{ourpaper},
which proposes the use of public-state additive, universal accumulators to realize a PKI and
is the only PKI system with a formal proof of security.  
Furthermore, our construction 
advances the state-of-the-art system of Patsonakis et al.~\cite{ourpaper} by eliminating the 
need for a trusted setup phase 
and, thus, leads to a truly decentralized PKI service. Lastly, the state
consumed by our construction is of constant size, contrary to systems such
as EthIKS, whose state grows linearly with the number of identities registered
in the system. We discuss the problems that arise from linear state (PKI)
constructions on public smart contract platforms
more thoroughly in Section~\ref{subsec:dummyeval}.

%

\section{Ethereum}
\label{sec:ethereum}

Ethereum is a blockchain-based platform for the development of smart contracts, i.e., stateful
agents that ``live'' in the blockchain and can execute arbitrary state transition functions.
Developing a smart contract 
involves writing its code in a high-level, Turing-complete programming language
(e.g., Solidity~\cite{solidity}), which is then compiled-down to Ethereum Virtual Machine (EVM)
initialization code. Contracts become part of Ethereum's global state (deployment) by wrapping their initialization code in a transaction, signing
it and broadcasting it to the network. The state and the code of smart contracts are
publicly accessible, thus, they can be trusted
for correctness, provided their code was properly audited and the blockchain is secure, but
not for privacy. Accessing the state of smart contracts can be performed
efficiently via Ethereum's light client protocol (\cite{ethlight}).
Users can interact with contracts by issuing transactions that specify the code to be executed
and its input arguments. In addition, contracts can call functions of other contracts,
which is known as a \emph{message call}, as a result of a user's transaction.
Ethereum's cryptocurrency is called \emph{ether} and serves as a means to incentivize participants
(miners) to engage in the protocol. 
Transactions fees, which compensate miners for their work, are expressed in a unit called
\emph{gas} and are a function of their byte size and the complexity of the code they invoke
(if any). Ethereum employs a flat cost model, i.e., each transaction byte and EVM operation
costs some predefined amount of gas (\cite{ethyellowpaper}). Transactions specify a \emph{gas
price}, which converts ether to gas and influences the incentive of miners to include it in
their next block. The higher the gas price, the higher its real monetary cost and priority
to be mined (\cite{ethereumgas}). A transaction that consumes $g_{cost}$ gas and specifies a
gas price of $g_{price}$ will cost $E = g_{cost} \times g_{price}$ units of ether.
Lastly, transactions and message calls, specify an upper bound on the amount of gas that they
can consume. This protects miners from, e.g., getting stuck in an infinite loop, an issue that
stems from Ethereum's Turing-completeness.

\section{Cryptographic Accumulators}
\label{sec:accumulators}

Cryptographic accumulators (first introduced in \cite{benalohacc}) provide a constant-sized
representation of a set of elements and allow for verifiable membership and, in some
cases, non membership queries. In the design of \cite{ourpaper}, the smart contract
acts as the accumulator manager and employs two instances of a public-state,
additive, universal accumulator with domain $\{0,1\}^*$.
First, as the smart contract's state and
code are publicly accesible, the accumulator's operations have to be performed by 
exclusively relying on public information (public-state). Second, the accumulator's
construction must, at least, support addition of elements (additive), to allow clients to
register their (identity,public-key) pairs. Third, universal accumulators allow for
\textbf{both} membership and non-membership verifications. Non membership
verifications are essential since clients interested in registering to the PKI have to
prove that their identity is not taken.
In accumulators, proofs of, e.g., (non) membership, are
referred to as \emph{witnesses}.

The relevant syntax of the public-state, additive, universal
accumulator of \cite{ourpaper} to the constructions implemented 
in this work is:

\begin{itemize}
\item $\mathsf{KeyGen}(1^\lambda)$: Generates a key pair $(pk,sk)$ and outputs $pk$.
 
\item $\mathsf{InitAcc}(pk)$: Outputs an accumulator value $c_0$, referring to the empty accumulated set $X\leftarrow\emptyset$.
 
\item $\mathsf{Add}(pk,x,c)$: Computes and outputs a new accumulator value $c'$ and $W$, 
a membership witness for $x\in X$. 

\item $\mathsf{MemWitGen}(pk,X,c)$: If $x\in X$, it outputs a membership witness $W$ for $x$.

\item $\mathsf{NonMemWitGen}(pk,X,c,x)$: If $x\notin X$, it outputs a non membership witness $W$ for $x$. 

\item $\mathsf{VerifyMem}(pk,x,W,c):$ If $W$ is an honestly produced membership witness for
$x\in X$, it outputs 1, otherwise, 0.

\item $\mathsf{VerifyNonMem}(pk,x,W,c)$: If $W$ is an honestly produced non membership
witness for $x\notin X$, it outputs 1, otherwise, 0.
\end{itemize}

Below, we provide information on the specific accumulator constructions
that we employ in our two implementations.

\textbf{RSA-based Universal Accumulator:}
The construction of the smart contract based PKI of Patsonakis et al.~\cite{ourpaper}
is built on top of the RSA-based universal accumulator of Li et al.~\cite{LiLX07}. 
This accumulator requires a trusted party
to run the $\KeyGen$ algorithm since, knowledge of the accumulator's secret key $sk$ can be used
to break its security.
To deal with the issue that the accumulator's input 
domain is restricted to prime numbers, the authors of \cite{ourpaper} incorporate
a deterministic procedure $\Map$ that, on input an arbitrary string, outputs a prime
number. This procedure is based on a function $f$, which is chosen uniformly at random
from a universal hash function family $U: \{0,1\}^{3k} \to \{0,1\}^k$ (\cite{CarterW79}). An arbitrary string $s$ is mapped to a prime number by,
first, computing the hash of the string, i.e., $h(s)\in\{0,1\}^k$, and
by repeatedly sampling from the set $\{x\in\{0,1\}^{3k}: f(x)=h(s)\}$ until a prime is found.
The sampling is performed by fixing the randomness to depend on the input string.  
After sampling $O(k^2)$ times, $\Map$ will output a prime number, except with negligible probability. 
To circumvent the fact that deletions in RSA accumulators require access to the accumulator's secret
key, the authors employ a
trick that is presented in \cite{BaldimtsiCDLRSY17}. Essentially, the $i$-th time an input
string $s$ is added or deleted, an element $x=(s,i,a)$ or $x=(s,i,d)$ is accumulated,
respectively. Lastly, in the RSA accumulator, both the accumulator's value and its
witnesses have constant bit size.


\textbf{Hash tree-based Universal Accumulator:}
The accumulator of Camacho et al.~\cite{CamachoHKO08}, which our construction employs, is
a public-state, additive, universal accumulator with the following 
differences and additional features. First and foremost, the accumulator is \textit{strong},
i.e., the accumulator manager is not required to be trusted.
Informally, a strong  accumulator does not 
require a trusted setup (there is no $\KeyGen$ algorithm). Second, it supports the deletion
of elements without relying on secret information. Third, it allows for additions/deletions
which are publicly verifiable. The latter is accomplished by an algorithm $\mathsf{CheckUpdate}$,
which, on input a witness returned by either $\Add$ or $\Delete$ and the accumulator's values before and after the update, it outputs 1, if the update was performed
honestly and 0, otherwise. The accumulator of \cite{CamachoHKO08} is based on collision-resistant hash functions and 
its underlying data structure is a balanced, binary, hash tree. In a few words, the 
accumulator's value is the hash of the root node and witnesses are hash path(s) starting from some node(s) (not necessarily leaf node(s)) that lead all the way up to the
root node. Note that the accumulator's value is of constant size, however, (non) membership and update witnesses have $\mathcal{O}(\lambda \log(n))$ bit size, where $n$
is the number of accumulated elements and $\lambda$ is the security parameter.

\section{Constructions}
\label{sec:constructions}

In this section, we illustrate how we employ the constructs of Sections
\ref{sec:ethereum} and \ref{sec:accumulators} to realize two versions of the smart contract
based PKI of \cite{ourpaper}. The first is the construction provided in \cite{ourpaper} and is based on
the RSA accumulator that was highlighted in Section~\ref{sec:accumulators}. The second, which we present in this work, is based on the Hash-tree accumulator of  \cite{CamachoHKO08}. 
The core idea of both schemes is decoupling the storage of
(identity,public-key) pairs from the verification of their validity.

The storage of information relevant to the protocol, e.g., (identity,public-keys) pairs, is offloaded to
an external database component. In \cite{ourpaper}, this component is modeled as an unreliable database functionality 
$\mathcal{F}_{UDB}$ (referred to as $UDB$ from now on), since the adversary is allowed to arbitrarily modify its state. As proven in \cite{ourpaper}, the scheme's security is not affected by such adversarial behaviour.  
The involvement of $UDB$ in the protocol is twofold. First, clients query $UDB$ to obtain information that will
allow them to, subsequently, interact with the smart contract. Second, following an
interaction with the smart contract, the clients post to $UDB$ information that reflect the
system's updated state. 


The smart contract maintains two cryptographic accumulators to facilitate the verification of 
the validity of identities, or, (identity,public-key) pairs. The first accumulator, $c_1$, 
accumulates (identity,public-key) pairs, allowing clients to infer if a pair is currently registered or not. 
The second accumulator, $c_2$, accumulates identities, allowing clients to infer if an
identity is registered or not. As shown in~\cite{ourpaper}, one accumulator
would suffice, however, at the expense of complicating the protocol's
presentation and increasing its computational complexity, issues we also want
to avoid here.
%
%
%



In both schemes, the smart contract is the most essential and expensive component to interact with in the system.  
Since its active involvement is required only during registration and revocation of (identity,public-key) pairs and due to lack of space, in this work, we only explain
how these two operations are performed. To register her pair, a client queries the $UDB$
to obtain the history of operations, which will allow her to compute a proof that her
identity is not accumulated in $c_2$. To revoke her pair, the client, instead, computes
a proof that her pair is accumulated in $c_1$ and proves possession of the
corresponding secret key. The latter is implemented via a digital signature on the public
key.

\subsection{RSA-based PKI}
\label{subsec:rsadpki}

This construction is based on the RSA accumulator, which
we highlighted in Section \ref{sec:accumulators}. Figure~\ref{rsapseudocode} of Appendix~\ref{sec:appendix}
illustrates the smart contract's pseudocode for this implementation. Here, $c_1$ accumulates $(id,pk,i,op)$
tuples and $c_2$ accumulates $(id,i,op)$
tuples, where $op=a$ or $op=d$.

A client that wishes to register her (identity,public-key) pair produces, at most, two
witnesses. First, a non membership witness $W_1$ for the tuple $(id,i,a)$ in
$c_2$. Second, and only if her identity
has been registered at least once in the past ($i \geq 2$), a membership
witness $W_2$ for the tuple $(id,i-1,d)$ in $c_2$.
Assuming both conditions hold, she will be able to
convince the smart contract that her identity is available. To construct these witnesses,
the client queries $UDB$ for the history of 
operations and locates records (if any) pertaining to her identity to find the 
proper value for index $i$. Then, she invokes the $\Register$
function of the smart contract and, as a result, will receive the updated values of
the accumulators and two new witness values, $W_1$ and $W_2$. These are membership
witnesses for the tuples $(id,pk,i,a)$ in $c_1$ and $(id,i,a)$ in $c_2$,
respectively. Next, the client computes a non membership witness $W_3$ for the tuple
$(id,i,d)$ in $c_2$. Lastly, she posts a $(\Register,id,pk,i,W_1,W_2,W_3)$
record to the $UDB$ which, among others, facilitates queries from other clients for
the validity of her mapping.

To revoke her (identity,public-key) pair, a client generates the following
proofs. First, a signature of her public-key ($\sigma_{sk}(pk)$). Second, a
membership witness $W_1$ for the tuple
$(id,pk,i,a)$ in $c_1$. Third, a non membership witness $W_2$ for the tuple
$(id,pk,i,d)$ in $c_1$. The witnesses are constructed similarly to
the case of registration, i.e., by querying $UDB$. Following a successful
revocation of her mapping, the client posts a $(\Revoke,id,pk,i)$ record to $UDB$.

\subsection{Hash tree-based PKI}
\label{subsec:hashdpki}

In this section, we propose an alternative construction, which is based on the Hash-tree accumulator of \cite{CamachoHKO08}. 
This accumulator supports additions and deletions which are publicly
verifiable. Thus, clients can perform, locally, additions and deletions of elements and
supply the smart contract with appropriate witnesses which prove that the operations were
performed honestly. To generate all involved witnesses, clients query $UDB$
for the history of operations.
Figure~\ref{hashpseudocode} of Appendix~\ref{sec:appendix} illustrates the smart
contract's pseudocode for this implementation.


To register an (identity,public-key) pair, the client generates a non membership
witness $W_2$ for her identity in $c_2$. She then performs, locally, the 
following updates. First, she accumulates the tuple $(id,pk)$ in $c_1$, which
produces an addition update witness $W_{add_{1}}$ and the updated value of the
accumulator $c_{add_{1}}$. Second, she accumulates her identity in $c_2$, which
produces an addition update witness $W_{add_{2}}$ and the accumulator's updated
value $c_{add_{2}}$. Assuming all values were computed honestly, the contract
will validate the proofs by invoking $\CheckUpdate(c_2,c_{add_{2}},W_{add_{2}},id)$, $\CheckUpdate(c_1,c_{add_{1}},W_{add_{1}},(id,pk))$ and $\VerifyNonMem\\(c_2,W_2,id)$, and update its accumulator
values. Lastly, the client posts a $(\Register,id,pk)$ to the $UDB$.

To revoke an (identity,public-key) pair, the client first signs her public-key 
($\sigma_{sk}(pk)$). Second, she generates a membership witness $W_1$ for 
the tuple $(id,pk)$ 
in $c_1$. She then performs, locally, two updates.
First, she deletes
her identity from $c_2$, which produces a deletion update witness $W_{del_{2}}$ and
the accumulator's updated value $c_{del_{2}}$. Second, she deletes the tuple
$(id,pk)$ from $c_1$, which produces a deletion update witness $W_{del_{1}}$ and
the accumulator's updated value $c_{del_{1}}$. Assuming all values were computed honestly, the contract
will validate the proofs by invoking $\CheckUpdate(c_1,c_{del_{1}},W_{del_{1}},(id,pk))$,
$\CheckUpdate(c_2,c_{del_{2}},W_{del_{2}},id)$ and $\VerifyMem(c_1,W_1,(id\\,pk))$, and update its accumulator
values. Lastly, the client posts a $(\Revoke,id,pk)$ record to the $UDB$,
which is not stored, but simply leads to the deletion of her registration record.

\section{Evaluation}
\label{sec:evaluation}

In this section, we present experiments that measure the cost of
running on Ethereum the 
constructions of Section~\ref{sec:constructions}, as well as
their building blocks.  Throughout this section, we intersperse our results with
recommendations for modifications and/or improvements to Ethereum that, we believe, 
are vital if Ethereum (or any smart contract platform) is to reach its maximum
potential of supporting arbitrary distributed applications (especially in the large scale~\cite{ethereumscale}).
We create a private blockchain that is
maintained by a single mining node. This eliminates the waiting time that transactions would
have in either the live or the test chain to be mined into a block. Thus, we are able to
measure accurately transaction gas costs and perform experiments on a larger scale.
We run our experiments on a CentOS 7 server that is equipped with an 8-core, 
64-bit, Intel(R) Xeon(R) CPU E5-2620 v4 @ 2.10GHz (with hyperthreading) and 32 GB
of RAM. We use the latest, stable release of \emph{geth} (v.1.8.17, \cite{geth}), the
official Ethereum client.
We conduct
our experiments via the \emph{truffle} suite (v.4.1.13, \cite{truffle}), a testing framework
that automatically handles compilation and deployment of contracts
and provides easy-to-use, JavaScript-based means of interacting with them. 
Finally, we use randomly generated 32-byte identities.

Our implementations employ a variety of primitives, e.g., signatures and accumulators.
One option would be to deploy each primitive as a separate library and have the
front-end PKI contract issue appropriate message calls.  Unfortunately, this option is the most
expensive in terms of
gas due to the extra cost of message calls (700 gas) and the increased
cost of reading the deployment address(es) of the library contract(s) from storage. 
The more efficient option is to pack all back-end logic into a single library and link it with the
front-end PKI contract. This eliminates the aformentioned costs. 
Thus, Ethereum imposes the following tradeoff. On the one hand, developers will tend
to pick the second option, as one of their main incentives is to minimize
gas cost. On the other hand, the first option: 1) promotes modular programming, 2) leads
to the construction of an on-blockchain ``standard library'', 
similar to what common programming languages have and most importantly, 3) mitigates duplicate logic, i.e.,
excess, duplicate state and code in the blockchain. Thus, reducing the costs of the
first option will aid in the development of future applications and incentivize
developers to adopt more modular programming approaches.

\textbf{Recommendation \#1: Significantly reduce the cost of issuing
message calls to libraries.}

Our first experiment provides insight regarding the overhead of a library
implementation, compared to a precompiled contract. Precompiled contracts
reside on well-known, static
addresses and require less gas because their code does not run in EVM assembly, but in
machine language of the physical node hosting the miner. 
We evaluate the cost of verifying 1,000 Secp256k1 elliptic curve signatures, based on
the library contract of \cite{secp256k1lib}. We measure a mean cost of 827,765.53 gas, 
with a standard deviation of 6,021.64 gas. 
At the time of this writing, the average \emph{gas limit} of blocks is about 8 million gas
(\cite{gaslimit}).  Thus, signature verification on the library contract
consumes $10.3\%$ of the current block gas limit, which is substantial.
In contrast, Ethereum's \emph{ecrecover}
precompiled contract, which operates on the same curve, costs only 3,000 gas.
Thus, the cost of the library implementation is two orders of magnitude higher,
which illustrates the benefits and 
importance of having built-in support for a variety of cryptographic operations.
In the evaluation of all the constructions that follow, we have modified the library
contract of \cite{secp256k1lib} to operate on the Secp256r1 curve. We repeat
the same experiment and measure the mean cost of signature verification to be
1,257,103.26 gas, with a standard deviation of 9,178.44 gas.

\subsection{RSA-based PKI Evaluation}
\label{subsec:rsaeval}

\begin{table*}[!t]
    \begin{center}
        \begin{tabular}{ | r | r | r | r | r |}
        \hline
        \multirow{2}{*}{ Operation } & \multicolumn{4}{  c | }{ Gas Cost } \\ \cline{2-5}
         & Min & Max & Mean & Std \\ \hline
        Mod. Mul. & 179,556 & 182,900 & 181,470.76 & 639.16 \\ \hline
        Mod. Exp. & 678,074 & 745,517 & 741,846.48 & 5,001.49 \\ \hline
        Primality Test & 1,481,160 & 1,502,502 & 1,490,219.13 & 5,480.23 \\ \hline
        $\Add$ & 810,030 & 810,158 & 810,153.32 & 17.61 \\ \hline
        $\VerifyMem$ & 755,130 & 755,796 & 755,537.23 & 126.63 \\ \hline
        $\VerifyNonMem$ & 1,473,345 & 1,525,685 & 1,519,279.96 & 5,386.87 \\ \hline 
        $\Map$ & 1,733,124,331 & 2,550,435,741 & 2,141,780,036 & 577,926,440.35 \\ \hline
        $\Register$ ($i=1$) & 89,801,425 & 8,620,016,945 & 1,681,994,990 & 1,313,539,096.67 \\ \hline
        $\Register$ ($i \geq 2$) & 89,160,026 & 10,676,126,282 & 2,575,538,734.5 & 1,715,254,997.91 \\ \hline
        $\Revoke$ & 440,467,878 & 13,805,874,517 & 3,598,585,618 & 1,910,918,965.1 \\ \hline         
        \end{tabular}
        \caption{Min, max, mean and standard deviation (columns 2-5) of the
        gas cost of: 1) 10,000 modulo multiplications, exponentiations and
        primality tests in the Big Number library, 2) 10,000 accumulations of primes
        ($\Add$) and (non) membership witness verifications ($\VerifyMem,\VerifyNonMem$)
        3) 1,000 mappings ($\Map$) of strings to primes and, 4) registrations 
        ($\Register$, for $(i=1)$ and $(i \geq 2)$) and revocations ($\Revoke$) of
        1,000 (identity,public-key) pairs in the RSA-based PKI.}
        \label{table:rsabuildingblocks}
    \end{center}

\end{table*}

In this section, we evaluate the RSA-based PKI of Section~\ref{subsec:rsadpki}, 
which employs the
following constructs. First, signature verification
via the Secp256r1 library contract. Second, arbitrary precision integer arithmetic, based
on the Big Number library developed by the Zerocoin team (\cite{bignumlib}). This library
supports operations which are relevant to this construction, such as modulo exponentiation  
and the Miller-Rabin probabilistic primality test. We
modify the implementation of the primality test because: 1) the original supports only 
a range of integers, whilst, our implementation supports all integers, 2) 
the original algorithm is seeded by externally provided randomness, which we modify to
be based on the hash of the last block, thus, limiting the adversary's knowledge
and influence on its output and, 3) the original does not perform 
sufficient iterations, which we modify to 
comply to the NIST standard (\cite{nistdss}), i.e., 64 witness loop iterations, thus,
the probability a composite number will be declared as prime
is $2^{-128}$. The third employed construct is the RSA
accumulator, which encompasses the $\Map$ procedure. 
The RSA moduli of the accumulators are 3072 bits long, thus, they
provide 128-bit security (\cite{bitsecurity}). Recall that $\Map$ uses a function $f:\{0,1\}^{3k}\rightarrow\{0,1\}^k$ chosen uniformly at random from a universal hash function family $U$. We set $\Map$'s parameter to $k=65$, i.e.,
$\Map$ outputs $3k=195$ bit primes. Thus, except for $1/2^{65}$ fraction of functions $f\in U$, a string will be mapped to a prime number, except with negligible probability (\cite{GennaroHR99}), which we deem reasonable.

We conduct four sets of experiments where the bit lengths of
the exponents, moduli and exponentiation bases
are 195, 3072 and 3072 bits long, respectively. First, we evaluate the operations
of the Big Number library that are relevant to this construction by running
10,000 
primality tests, modulo multiplications and exponentiations, respectively. Second, we evaluate
the RSA accumulator by running 10,000 iterations of each of the following operations: 1)
accumulations ($\Add$) of 195-bit prime numbers, and 2) (non) membership witness
verifications ($\VerifyMem,\VerifyNonMem$). Third, we
measure the cost of 1,000 mappings ($\Map$) of strings to 195-bit
prime numbers. Fourth, we measure the cost of registering
($\Register$) and revoking ($\Revoke$) 1,000 (identity,public-key) pairs in
the RSA-based PKI. Recall that registration differentiates between two cases,
i.e., whether an identity is registered for the first time ($i=1$), or not 
($i \geq 2$). Table~\ref{table:rsabuildingblocks} illustrates the results.

Regarding the Big Number library experiments,  Table~\ref{table:rsabuildingblocks} shows
that modulo 
exponentiation and primality testing are the more expensive operations. The former
is based on one of Ethereum's precompiled contracts (\cite{modexpeth}) and its cost
is dominated by the exponent's length, especially in cases where it is larger
than 32 bytes. In addition, this operation is invoked in the (main) witness loop of the Miller-Rabin
test, thus, the data suggest that, on average, the primality test performs two loop iterations.
To compute the cost of the RSA accumulator's operations, we have to factor in the cost of 
reading from storage the accumulator's value, its exponentiation base and its modulus
(a total of 36 EVM words). The cost of reading an EVM word (32 bytes) from
storage is 200 gas, thus, $36 \times 200 = 7,200$ extra gas. The 
$\VerifyMem$ operation involves one modulo exponentiation. The $\VerifyNonMem$ operation
involves two modulo exponentiations and one modulo multiplication. Thus, as the data
suggest, the gas cost of these two operations follows directly from that of reading
the appropriate values from the contract's storage and the invoked operations of the
Big Number library.
The $\Add$ operation involves one modulo exponentiation and modifies the accumulator's value. Thus, in addition to the 
aforementioned cost of reading from the contract's storage, there is also
the cost of storing the accumulator's updated value to the contract's state.
Updating an EVM word on storage costs 5,000 gas. The 
accumulator is 12 EVM words long, thus, $12 \times 5,000 = 60,000$ extra gas, or, a total
of $60,000 + 7,200 = 67,200$ gas.

The key result of this section is our demonstration of
the practical implications of the RSA-based PKI's design. Recall that in this construction,
the smart contract's state and (non) membership witnesses have constant
size at the expense of computational overhead. This tradeoff is embodied by the $O(k^2)$
complexity of the
$\Map$ procedure, which is involved in both the $\Register$ and $\Revoke$ operations of 
the RSA-based PKI, to map the contract's inputs to prime numbers. These prime numbers are
then input to the appropriate witness verification algorithms of the RSA
accumulator. For instance, the $\Revoke$ operation involves one signature verification, one invocation of $\VerifyNonMem$ and $\VerifyMem$ each and three invocations of $\Map$. We
have already illustrated that the cost of signature verification and of the RSA accumulator's
operations is deterministic, however, the same cannot be said about $\Map$.
While $\Map$ is
deterministic in terms of its output, the number of iterations it performs to produce its output is not.
Thus, we can have cases where one invocation of $\Map$ costs more than a $\Register$
or $\Revoke$ operation of the RSA-based PKI, as illustrated by the heavily skewed data 
of Table~\ref{table:rsabuildingblocks}. Consequently, $\Map$ dominates the cost of the
RSA-based PKI's operations. We perform an additional experiment where
we measure the cost of running \textbf{one iteration} of $\Map$ on input of 100,000 strings.
We measure the average gas cost to be 9,488,542.32 gas, with a standard deviation of 
17,794.86 gas. Consequently, even one iteration of $\Map$ exceeds Ethereum's block gas limit.

\textbf{Result \#1: The provably secure, RSA-based smart contract
PKI is not viable in Ethereum.}

\textbf{Discussion:} There are two reasons why $\Map$'s gas cost is so high. First,
in Ethereum, it is cheaper to access one EVM word (32 bytes) than one byte. This ``was chosen to facilitate the Keccak256
hash scheme and elliptic-curve computations'', as stated in Ethereum's yellow
paper (\cite{ethyellowpaper}). It has nothing to do with efficiency as no real-world
physical machine, on top of which the EVM runs,
supports 32 byte words. Therefore, one potential improvement would be to modify Ethereum's
cost model to account for this contradiction. For instance, accessing a single byte could
be simply tuned to $1 \over 32$ of the cost of loading an EVM word. This 
 change, apart from being more fair, allows for more packed data encodings which
can reduce the size of  transactions and, as a result, the size of the blockchain.
Second, $\Map$'s computation revolves around bit operations, which are, currently, very
expensive in Ethereum as they have to be performed via the EVM's integer exponentiation
function. For instance, setting one bit of a memory byte array costs 586 gas. However, 
in the near future, the EVM will support bitwise shifting (\cite{ethbitwiseops}), which will
only cost 3 gas and, thus, will provide substantial improvements for $\Map$.

\textbf{Recommendation \#2: Ethereum's cost model should be modified to account for the
granularity of the data that are accessed.}

Nevertheless, for the RSA-based PKI, there are other
alternatives that we can explore, which we leave as future work, that will also benefit
from all the aforementioned propositions, such as verifiable computation
(\cite{verifiablecomputation}). In this setting, the smart contract will need only to
verify proofs that the client computed $\Map$ correctly, instead of invoking
$\Map$ itself.

\subsection{Hash tree-based PKI Evaluation}
\label{subsec:hasheval}

\begin{figure*}
\centering
    \begin{subfigure}{.49\textwidth}
        \includegraphics[width=.7\linewidth,angle=270]{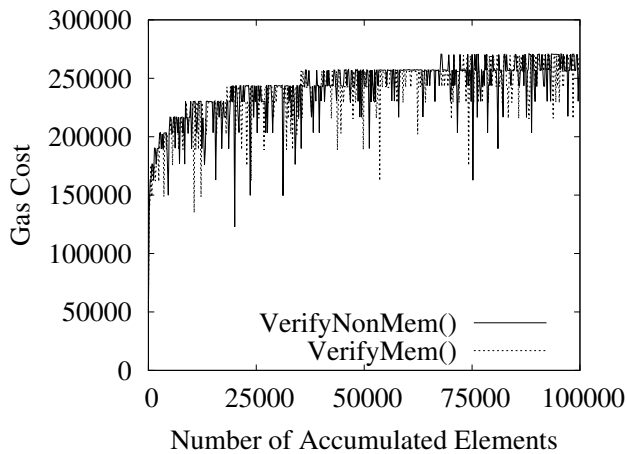} 
        \caption{}
        \label{subfig:hashaccmemnonmemver}
    \end{subfigure}
    \begin{subfigure}{.49\textwidth}
        \includegraphics[width=.7\linewidth,angle=270]{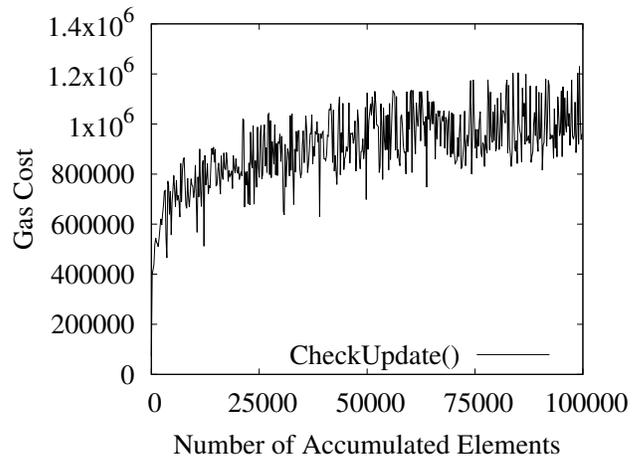}
        \caption{}
        \label{subfig:hashacccheckupdate}
    \end{subfigure}  
\caption{Gas cost versus the number of accumulated values of 100,000 of the
following hash tree accumulator verifications: (\subref{subfig:hashaccmemnonmemver}) (Non) Membership
witnesses and, (\subref{subfig:hashacccheckupdate}) Accumulator updates.
Note the different y-axis scales.}
\label{fig:hashacc}
\end{figure*}

\begin{figure}
        \includegraphics[width=.7\linewidth,angle=270]{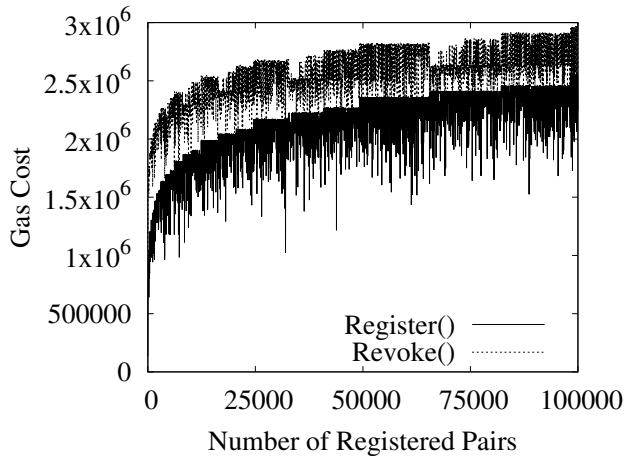} 
\caption{Gas cost versus the number of registered (identity,public-key)
pairs of the registration and revocation operations of 100,000 pairs in the
Hash-based PKI.}
\label{fig:hashdpki}
\end{figure}

In this section, we evaluate the Hash-based PKI, which we
proposed in Section~\ref{subsec:hashdpki}. This construction employs the following
primitives: 1) the Secp256r1 library contract for signature verification
and, 2) the Hash tree accumulator of
Section~\ref{sec:accumulators}. 
We employ the SHA-256 hash function, which is exposed as a precompiled contract
in Ethereum. 
Recall that in the Hash-based PKI, all involved proofs are a logarithmic function
of the number of accumulated elements, in contrast to the RSA-based construction, where
they are constant-sized. 


In our first experiment, we accumulate a total of 100,000 elements and measure the cost of
the Hash tree accumulator's operations. In Figures~\ref{subfig:hashaccmemnonmemver} and
\ref{subfig:hashacccheckupdate}, we plot the gas cost versus the number of
accumulated elements of verifying
(non) membership witnesses ($\VerifyMem,\VerifyNonMem$) and
accumulator updates ($\CheckUpdate$), respectively. The general trend
resembles, as expected, that of a logarithmic function. However, the curves have a large number
of dips. This is because in this scheme, proofs are paths of the
accumulator's hash tree. The size of each proof depends on the position of the starting
node(s) in the hash tree and, thus, its verification cost varies. As illustrated in the graphs,
verifying accumulator updates is more expensive than that of (non)
membership witnesses. Indeed, the size of the former proofs tends to be two,
or even, three times the size of the latter, which is reflected in their respective verification
costs.

In our second experiment, we evaluate our Hash-based PKI construction. In Figure~\ref{fig:hashdpki},
we plot the gas cost versus the number of registered pairs of registering
and revoking 100,000 (identity,public-key) pairs. The results show that revocation is the more
costly procedure as it involves the added cost of verifying signatures. The cost of the most expensive revocation that we measured was 2,999,214
gas, i.e., $37.4\%$ of the current block gas limit. Thus, in terms of gas cost, this construction can be deployed on the main chain of Ethereum. 

\textbf{Result \#2: Our Hash-based PKI construction is viable for deployment on the main chain of Ethereum.}

\textbf{Discussion:} The Hash-based PKI is best suited for small, to moderately sized PKIs, since the involved
proofs are not of constant size. 
Assuming that the number of registered pairs monotonically increases,
there will come a point where verifying proofs will exceed the block's gas limit. Recall that the
cost of a transaction is a function of its computational complexity \textbf{and} its
byte size. One
might argue that this issue can be balanced out by an increase in the block's gas limit, which
is certainly the observed trend up to the time of this writing (\cite{gaslimit}). However, a miner's main
incentive is to produce (hash) blocks as fast and with as low operational costs as possible. Thus, it can be expected  that the increase in the block's gas limit will, eventually, plateau.
This line of reasoning assumes
that the blockchain's consensus mechanism revolves around Proof-of-Work (PoW), as is 
currently the case. However, Ethereum is planning to replace PoW with
Proof-of-Stake (PoS), a consensus protocol that requires a small amount of computation.
Discussing PoS is out of the scope of this paper, however, it is reasonable to assume 
that it will change the incentives of miners. Indeed, in this computationally light paradigm, miners might be
willing to expend their computational resources to mine blocks that contain larger 
transactions to maximize their rewards. This will gradually increase the block's gas limit
which, as a result, will favor the scale of our Hash tree-based PKI even more.

We now illustrate a few important points regarding hash functions and precompiled contracts.
Ethereum supports three hash functions: 1) RIPEMD160, whose computation costs 600 gas, 
plus 120 gas per input word, 2) SHA-256, whose computation costs 60 gas,
plus 12 gas per input word, 3) KECCAK-256, whose computation costs 30 gas,
plus 6 gas per input word. Functions (1) and (2) comply with the NIST standard and 
are implemented as precompiled contracts, however, function (3), does not comply with the
standard and is implemented as an EVM opcode. These distinctions have interesting
implications. Because functions (1) and (2) are precompiled contracts, they incur the extra
gas cost of a message call (700 gas), while function (3) does not. Consequently, Ethereum's
cost model encourages the use of a non-standard-compliant
hash function. Thus, application developers are forced to either code the client side (at least
in part) in 
JavaScript, for the sole purpose of having access to Ethereum's non-standard implementation
of (3), or, pay the extra gas cost. As one of the main incentives of developers in these
platforms is to minimize gas costs, the aforementioned distinctions essentially \emph{encourage}
client implementations that are unnecessarily complicated and limit the use of standard,
mature and efficient libraries, such as \textit{libgcrypt} (\cite{libgcrypt}).
Recently, a proposal has been submitted (\cite{precompiledcall}) to address this issue. 
If accepted, 
this change will further diminish the gas cost of the Hash tree-based PKI, thus,
increasing its deployment scale. 

\textbf{Recommendation \#3: Reduce message call costs from precompiled contracts and equalize the costs of all supported hash functions.}

Lastly, as illustrated previously, Ethereum is inconsistent in the way it
handles and exposes more complicated instructions. Given that: 1) the size of EVM
assembly opcodes is one byte, 2) most values are already in use
(\cite{ethevmopcodes}) and, 3) the purpose of assembly language is to 
provide access
to low-level instructions, we believe that more sophisticated functionality (e.g., hash functions) should be offloaded
to a standard library of precompiled contracts. Furthermore, Ethereum should design and incorporate a more
developer-friendly way of addressing these contracts. Currently, developers need to memorize
(or look up) the address in which each contract resides, e.g., SHA-256 resides in address 0x02,
which is cumbersome. Convenient helper functions
in EVM assembly that are translated to the appropriate message calls would be helpful.

\textbf{Recommendation \#4: Sophisticated functionality should be moved to a standard library
of precompiled contracts that can be addressed in a developer-friendly manner.}

%
%
%

\subsection{Linear State PKI Evaluation}
\label{subsec:dummyeval}

In this section, we present experiments that evaluate a simple smart contract
PKI which stores all (identity,public-key) pairs in the contract's state. This is the same approach that prior proposals employ (e.g., \cite{ethens,scpki,bbpki}), including the Ethereum Name
Service,
and allows us to illustrate the shortcomings of Ethereum's
pricing of storage. In this simple scheme, registration and revocation are
straightforward 
processes. During registration, the contract checks if there is an entry for the 
input identity in its state. Assuming there isn't one, it adds it. During revocation, 
the contract first validates the input signature, as in the prior two constructions, and checks 
if there is an entry in its state for that identity. If so, the contract simply removes it 
from its state. Table~\ref{table:dummypki} illustrates the gas cost of registering and
revoking 10,000 (identity,public-key) pairs in this setting. During
revocation, part of the contract's storage is freed, and the transaction is refunded gas. As
a result,
the overall cost of revocation is less than the verification of a Secp256r1 signature.

\begin{table}
    \begin{center}
        \begin{tabular}{ | c | c | c | c | c|}
        \hline
        \multirow{2}{*}{ Operation } & \multicolumn{4}{  c | }{ Gas Cost } \\ \cline{2-5}
         & Min & Max & Mean & Std \\ \hline 
        $\Register$ & 89,469 & 89,661 & 89,643.91 & 33.74 \\ \hline
        $\Revoke$ & 904,197 & 949,505 & 931,150.28 & 6,410.29 \\ \hline     
        \end{tabular}
        \caption{Min, max, mean and standard deviation (columns 2-5) of the gas cost of registering and revoking 10,000 randomly generated (identity,public-key) pairs in the Linear State PKI contract.}
        \label{table:dummypki}
    \end{center}

\end{table}
\textbf{Discussion:} Clearly, this is, currently, the least costly approach to deploy
on Ethereum and the
reasons are straightforward.
First, it is light in terms of computation. Indeed, excluding
signature verification, which is the dominant cost of revocation,
the contract spans a total of 10 lines of Solidity
code, consisting solely of a few \emph{if} statements. Second, as stated by
Buterin (\cite{transactionfeeeco}), storage is
extremely underpriced and, as of yet, there is no incentive for freeing it. 
However, the issue of storage and its effect on the size of Ethereum's state
is complex.  
Miners decide whether they include a 
transaction in a block according to their ``private cost'', i.e., their 
own private resource expenditure. Regardless, their decision affects the entire
network, as all participating nodes have to download and validate newly
mined blocks, without having a choice in the matter. For instance, if we take the 
Linear State PKI as a reference point, it might be favorable for some miner(s) to store
a few bytes on disk, e.g., due to having abundant and inexpensive disk space. However,
that may not be the case for other nodes participating in the protocol.
This suggests a ``social cost'' of transactions which may not be completely aligned 
to an individual  miner's private costs. If this social cost of 
transactions is not completely accounted for, the increasing size of Ethereum's state may deincentivise 
new full nodes from entering the system. Furthermore, the size of Ethereum's
state can also be an obstacle to nodes
merely syncing with the system. This issue affects a variety of
topics ranging from, light clients (e.g., smartphones) being able to
interface with smart contracts, to blockchain security. 

One of the proposed
countermeasures is imposing small, \textbf{static} rent fees on contracts (\cite{ethereumrent}) so as to avoid 
being ``deactivated'', i.e., no one being able to interact with them. However, it is
Ethereum's \textbf{static} cost model that has caused this
problem in the first place. We believe storage is a special commodity and that its price
should be dynamically adjusted. Base storage
price should depend on the global size of Ethereum's state, i.e., the bigger
the size of its state, the higher the base storage price. In addition, the cost of transactions
that increase the size of a contract's state should scale accordingly, thus,
providing a counterincentive to over-utilizing contract storage.

\textbf{Recommendation \#5: Storage costs should take into account the current blockchain's size as well as the size of the invoked contract's state.}

Clearly, the issue of storage costs is complex and remains an open problem requiring future
research to be addressed properly.  Nonetheless, the constant state PKI constructions discussed in
this work are well aligned with
space optimal use of  smart contract platforms.

\bibliographystyle{IEEEtran}
\bibliography{IEEEabrv,bibliography}

\appendix

\section{Appendix}
\label{sec:appendix}

\begin{figure}[H]
\begin{framed}
\small
Smart Contract State: $N_1,N_2,k \in \mathbb{Z},c_1 \in \mathbb{Z}_{N_{1}},
g_1 \in QR_{N_{1}},c_2 \in \mathbb{Z}_{N_{2}},
g_2 \in QR_{N_{2}},A \in GF(2)^{k \times 3k}$

\begin{enumerate}

\item $\Constructor(g_1,N_1,g_2,N_2,k,A):$ 

Store input values $g_1,N_1,g_2,N_2,k,A$ to the corresponding state variables

$c_1 = \InitAcc(g_{1},N_{1})$

$c_2 = \InitAcc(g_{2},N_{2})$

\item $\Register(id,pk,i, W_1, W_2):$ 

$x_{p_{2}} \leftarrow \Map(k,A,(id,i,a))$

if $\VerifyNonMem(pk_2,x_{p_{2}},W_1,c_2)=0$
    
    \quad return $\fail$

endif

if $i\geq 2$

    \quad $x'_{p_{2}} \leftarrow \Map(k,A,(id,i-1,d))$
    
    \quad if $\VerifyMem(pk_2,x'_{p_{2}},W_{2},c_2)=0$
    
        \qquad return $\fail$
        
    \quad endif

endif

$x_{p_{1}} \leftarrow \Map(k,A,(id,pk,i,a))$

$(c_1,W_1)\leftarrow\Add(pk_1,x_{p_{1}},c_1)$

$(c_2,W_2)\leftarrow\Add(pk_2,x_{p_{2}},c_2)$

return $(c_1,W_1,c_2,W_2)$

\item $\Revoke(id,pk,i,W_{1},W_{2},\sigma_{sk}(pk)):$

$x_{p_{1}} \leftarrow \Map(k,A,(id,pk,i,d))$

$x'_{p_{1}} \leftarrow \Map(k,A,(id,pk,i,a))$

if $\VerifyNonMem(pk_1,x_{p_{1}},W_2,c_1)=0\vee \VerifyMem\\(pk_1,x'_{p_{1}},W_{1},c_1)=0 \vee \VerifySig(\sigma_{sk}(pk),pk)=0 $

    \quad return $\fail$
    
endif

$x_{p_{2}} \leftarrow \Map(k,A,(id,i,d))$

$(c_1,W_1)\leftarrow\Add(pk_1,x_{p_{1}},c_1)$

$(c_2,W_2)\leftarrow\Add(pk_2,x_{p_{2}},c_2)$

return $(c_1,W_1,c_2,W_2)$

\item $\RetrieveState():$ 

return $(c_1,c_2,k,A)$

\end{enumerate}
\end{framed}
\caption{Pseudocode of the smart contract of the RSA-based construction. The
$\Constructor$ function is executed once, during contract deployment and
initializes the contract's state. The remaining functions constitute the main interface of the
smart contract, i.e., registering ($\Register$) and revoking ($\Revoke$) an $(id,pk)$ pair 
and retrieving its current state ($\RetrieveState$).}
\label{rsapseudocode} 
\end{figure}

\begin{figure}[t!]
\begin{framed}
\small
Smart Contract State: $c_1,c_2,\lambda_1,\lambda_2 \in \mathbb{Z}$

\begin{enumerate}

\item $\Constructor(\lambda_1,\lambda_2):$ 

Store input values $\lambda_1,\lambda_2$ to the corresponding state variables

$c_1 \leftarrow \InitAcc(\lambda_1)$

$c_2 \leftarrow \InitAcc(\lambda_2)$

\item $\Register(id,pk,W_2,c_{add_{1}},W_{add_{1}},c_{add_{2}},W_{add_{2}}):$ 

if $\mathsf{sizeof}(id)\neq \lambda_2\vee\CheckUpdate(c_2,c_{add_{2}},W_{add_{2}},id)\\=0\vee \mathsf{sizeof}(id,pk)\neq \lambda_1 \vee \CheckUpdate(c_1,c_{add_{1}},\\W_{add_{1}},(id,pk))=0\vee\VerifyNonMem(c_2,W_2,id)=0$
    
    \quad return $\fail$

endif

$c_1 \leftarrow c_{add_{1}}$

$c_2 \leftarrow c_{add_{2}}$

\item $\Revoke(id,pk,W_1,\sigma_{sk}(pk),c_{del_{1}},W_{del_{1}},c_{del_{2}},W_{del_{2}}):$

if $\mathsf{sizeof}(id) \neq \lambda_2 \vee \mathsf{sizeof}(id,pk) \neq \lambda_1 \vee \VerifyMem(c_1,\\W_1,(id,pk))=0 \vee \VerifySig(\sigma_{sk}(pk),pk)=0 \vee \CheckUpdate(c_1,c_{del_{1}},W_{del_{1}},(id,pk))=0
\vee \CheckUpdate(c_2,c_{del_{2}},W_{del_{2}},id)=0$
    
    \quad return $\fail$

endif

$c_1 \leftarrow c_{del_{1}}$

$c_2 \leftarrow c_{del_{2}}$

\item $\RetrieveState():$ 

return $(c_1,c_2,\lambda_1,\lambda_2)$

\end{enumerate}
\end{framed}
\caption{Pseudocode of the smart contract in the Hash-based construction. The
$\Constructor$ function is executed once, during contract deployment and
initializes the contract's state. The remaining functions constitute the main 
interface of the smart contract, i.e., registering ($\Register$) and revoking ($\Revoke$)
an $(id,pk)$ pair and retrieving its current state ($\RetrieveState$). The $\mathsf{sizeof}$
operator outputs the number of bits of its input.}
\label{hashpseudocode} 
\end{figure}

In this section, we present figures that illustrate the pseudocode of the smart contract 
pertaining to each construction of Section~\ref{sec:constructions}.
In both cases, the interface of
the smart contract is comprised of four functions. The $\Constructor$ is invoked only
when the contract is deployed and serves as a means to initialize its state.
The $\Register$ and $\Revoke$ functions constitute the main operations of the smart
contract for registering and revoking, respectively, (identity,public-key) pairs. 
Lastly, $\RetrieveState$ is a helper function that clients invoke to obtain the current
state of the smart contract, e.g., via Ethereum's \emph{web3.js} API (\cite{ethweb3}).
The Solidity code of the smart contracts, as well as, that of all the building blocks that we
employ in our implementations, can be made available upon request.

Figure~\ref{rsapseudocode} illustrates the pseudocode of the RSA-based PKI of
Section~\ref{subsec:rsadpki}. Its state is comprised of the following: 1) two RSA
accumulator values, $c_1$ and $c_2$, 2) the accumulators' respective RSA moduli,
$N_1$ and $N_2$, 3) the accumulators' exponentiation bases, $g_1$ and $g_2$, 
which are quadratic residues modulo $N_1$ and $N_2$, respectively and, 4) the
parameters of the $\Map$ procedure where, $k$, is its security parameter that, among others,
determines the bit length of the primes that $\Map$ outputs ($3k$ bits long) and, $A$, is a
randomly generated $k \times 3k$ bit matrix.

Figure~\ref{hashpseudocode} illustrates the pseudocode of the Hash-based PKI of
Section~\ref{subsec:hashdpki}. Its state is comprised of the following: 1) two Hash
accumulator values, $c_1$ and $c_2$ and, 2) the accumulators' respective security
parameters, $\lambda_1$ and $\lambda_2$, which, essentially, determine the number of bits of
the accumulators' inputs. To this end, and as illustrated in the smart contract's
pseudocode, we employ a $\mathsf{sizeof}$ operator which, on input an arbitrary string $s$,
outputs its bit length.

\end{document}